# Direct Observation and Control of Surface Termination in Perovskite Oxide Heterostructures

*Thomas Orvis[⸙], Tengfei Cao[†], Mythili Surendran[⸙], Harish Kumarasubramanian, Austin Cunniff, Rohan Mishra[†], and Jayakanth Ravichandran[δ]**

Mork Family Department of Chemical Engineering and Materials Science, University of Southern California, 925 Bloom Walk, Los Angeles, CA 90089, USA

**ABSTRACT**

The interfacial behavior of quantum materials leads to emergent phenomena such as two dimensional electron gases, quantum phase transitions, and metastable functional phases. Probes for in situ and real time surface sensitive characterization are critical for active monitoring and control of epitaxial synthesis, and hence the atomic-scale engineering of heterostructures and superlattices. Termination switching, especially as an interfacial process in ternary complex oxides, has been studied using a variety of probes, often ex situ; however, direct observation of this phenomena is lacking. To address this need, we establish in situ and real time reflection high energy electron diffraction and Auger electron spectroscopy for pulsed laser deposition, which provide structural and compositional information of the surface during film deposition. Using this unique capability, we show, for the first time, the direct observation and control of surface termination in complex oxide heterostructures of SrTiO3 and SrRuO3. Density-functional-theory calculations capture the energetics and stability of the observed structures and elucidate their electronic behavior. This demonstration opens up a novel approach to monitor and control the composition of materials at the atomic scale to enable next-generation heterostructures for control over emergent phenomena, as well as electronics, photonics, and energy applications.

## INTRODUCTION

Complex oxide heterostructures are an active area of research which address fundamental questions in solid-state systems,[1-5] and have broad technological applications[6-10]. Dramatic advances in the techniques for the growth and *in situ* structural characterization of complex oxides has led to atomistic understanding of growth processes and the discovery of emergent physical and chemical phenomena.[11-14] Despite these achievements, a detailed understanding of bulk and interfacial composition in these heterostructures still remains challenging.[14,15] Limits on compositional probes are the bottleneck to preparing thin film heterostructures with well-defined interfaces.[16-18] Hence, advances in *in situ* and real time elemental and chemical composition analysis would significantly improve our ability to prepare atomically precise complex oxide heterostructures with real time feedback control.

Electron diffraction methods, such as reflection high energy electron diffraction (RHEED) and low energy electron diffraction (LEED), have been used for *in situ* structural characterization in thin film deposition systems since as early as the 1960's,[19] but were not associated with real time growth-rate observation until 1981.[20] As the *in situ* and real time application of RHEED improved,[21] with modifications for compatibility with higher pressure oxide deposition systems,[22] the quality and precision of the synthesized complex oxide thin films and heterostructures utilizing growth methods, such as pulsed laser deposition (PLD), improved dramatically.[11] Compared to structural analysis, elemental and chemical composition analysis is significantly more challenging due to the need for judicious use of standards and a high vacuum environment for reliable data acquisition. Furthermore, the harsh high pressure and oxidizing environment of typical complex oxide deposition systems like PLD poses a significant technical challenge to this task. One of many composition analysis methods, Auger electron spectroscopy (AES) has been used as a powerful surface characterization technique for over half a century,[23] yet only recently has the development of probe design allowed *in situ* and real time characterization of complex oxide surfaces during their deposition with this technique.[24] Recently, AES was shown as a facile method to characterize *in situ* atomic-scale elemental composition with pulsed laser deposition.[25] Here, we show that AES can not only be used for *in situ* and real time studies of elemental composition analysis, but the technique can unambiguously identify subtle phenomena, such as surface termination switching in oxide heterostructures, which has been under considerable debate[26-29].

Surface and interface termination can have a dramatic impact on the physical and chemical properties of epitaxial thin film heterostructures. For example, the electronic properties of nitride electronics and photonics are highly dependent on their termination (Ga-polar or N-polar) and orientation.[30,31] Similarly, complex oxide heterostructures, such as $LaAlO_3/SrTiO_3$ show conducting ($TiO_2$-LaO interface) and insulating (SrO-$AlO_2$) interfaces depending on interfacial termination;[12] an observation which led to a flurry of studies showing that interface and surface termination influence a broad range of properties. Controlling surface termination of complex oxides synthesized with PLD, however, has been limited to approaches such as substrate preparation and of the deposition of $SrRuO_3$ to obtain an SrO-terminated surface.[26] Nevertheless, there are very few reports that probe surface or interface termination of $LaAlO_3/SrTiO_3$ heterostructures during the growth process due to *in situ* characterization limitations, and *ex situ* results may not fully represent the growth dynamics present.[27] The inability to study surface termination has caused wide variability in results reported on complex oxide heterostructures and has stifled the community's efforts to fully understand the influence of termination on their properties.[29,32,33]

In this work, the surface termination of a prototypical perovskite oxide $SrTiO_3$ (STO) is probed quantitatively and then deliberately switched, *in situ* and real time, during the growth process. The composition evolution monitored with *in situ* AES is not only sensitive to these termination switching events, when combined with RHEED, it also provides the precise stopping points for synthesizing structurally and chemically exact heterostructures. We establish the atomic-monolayer sensitivity of the AES technique by using a parameter-free Auger electron escape depth model to quantitatively compare the experimentally measured Auger signal intensity. Furthermore, the surface energies of SrO-, $TiO_2$-, and $RuO_2$-terminated STO and $SrRuO_3$ in bulk and thin film states as well as in $SrRuO_3$/STO heterostructures are analyzed based on density-function-theory (DFT) calculations, and applied to explain the energetic origins of the $RuO_2$- to SrO-termination conversion observed with the deposition of $SrRuO_3$. These results show our ability to study characteristics of thin film growth that were previously speculated upon but never directly observed, such as termination switching events and dynamic layer-rearrangement, which occur on a timescale comparable to the deposition rate.[26-37] This demonstration dramatically broadens the knowledge of atomic scale composition and provides unprecedented control over the quality of complex oxide heterostructures which can be synthesized using epitaxial growth techniques such as PLD.

## RESULTS AND DISCUSSION

Complex oxide thin films were grown on (001) single-crystal $TiO_2$-terminated STO substrates by PLD with *in situ* RHEED and AES. A detailed description of growth methods and parameters can be found in the Experimental Section. The plume generated by the pulsed laser ablation of the target material interferes with the Auger probe's ability to collect spectra, making reliable acquisition possible only between laser pulses. Lengthy pauses up to 30 seconds between depositions, required to accommodate the collection of spectra, can result in ample time for the recovery of the specular spot intensity in RHEED, suggesting surface smoothing. While this may promote the quality of the growth, as has been reported, [38] it can interfere with the observation of dynamic growth events. One solution is to increase the sampling rate, which also requires improving the quality of the signal. Therefore, we developed a pulse-probe technique which utilizes the rapid collection of narrow-energy spectra between pulsed laser bursts. To improve the signal quality, we have selected low energy transitions for observation when possible, which has the simultaneous benefit of improving the surface-sensitivity of the probe due to the energy-dependence of the inelastic mean free path.[39]

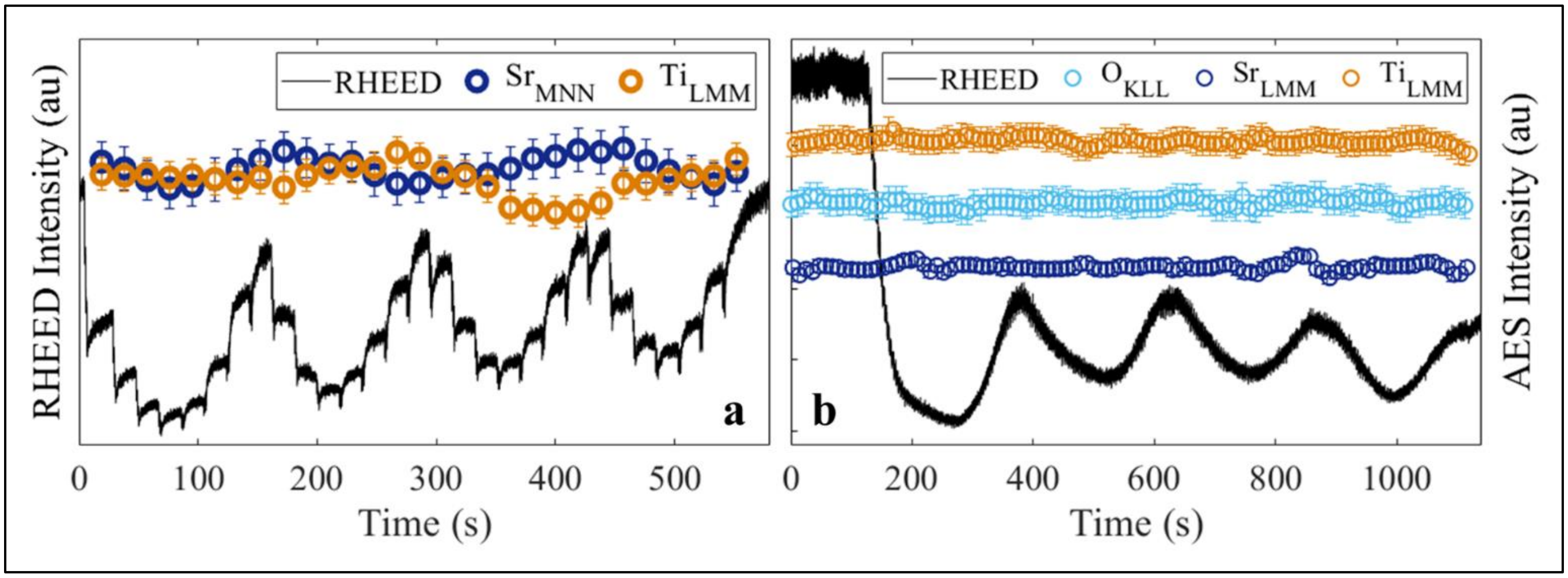

**Figure 1.** Measured intensity of RHEED specular spot (noted as RHEED intensity) and AES data simultaneously collected using the pulse-probe method outlined in the text for the growth of homoepitaxial STO. The Auger lines used to monitor Sr and Ti are $Sr_{MNN}$, and $Ti_{LMM}$ transitions in (a), and to monitor O, Sr and Ti are $O_{KLL}$, $Sr_{MNN}$, and $Ti_{LMM}$ transitions in (b). Laser pulses result in a roughening of the surface (a rapid decrease in RHEED intensity) followed by recovery during Auger acquisition (gradual increase in RHEED intensity). Depending on the duration of the spectral collection and the number of pulses between scans, the shape of the resulting oscillations will vary with either long scans and long recovery time as shown in (a), or short scans and thus short recovery as shown in (b). Likewise, the sensitivity of the scans to compositional changes is dependent on the number of data points and their integration time, resulting in a trade-off between the quality of RHEED and AES data.

Using this pulse-probe technique, we show real time *in situ* observation of surface composition in homoepitaxial STO, using both slow and fast Auger spectra acquisition rates, shown as simultaneous Auger and RHEED intensity in **Figure 1**. This growth approach is critical to successfully deploying real time and *in situ* techniques for direct characterization of surfaces during thin film deposition. We note that the Auger signal intensity is independent of the surface roughness as there is no correlation between RHEED specular spot intensity and AES intensity, and the AES intensities of the individual elements remain relatively constant throughout the growth, indicating no significant composition evolution during the growth. With these factors in mind, we will show that this technique is sufficiently sensitive to witness subtle compositional evolution demonstrated in later results and the Supplementary Information. Additionally, the pulse-probe technique shows sufficient temporal sensitivity to monitor subtleties of dynamic complex oxide thin film growth events never witnessed before with previously demonstrated *in situ* characterization methods.[40-42]

We chose to study surface termination in model heterostructures of $SrRuO_3$ and $SrTiO_3$ to evaluate the suitability of the AES technique to monitor subtle surface composition changes. To demonstrate the surface-sensitivity of our technique, we synthesized a film with the following structure: two unit cells of homoepitaxial STO on a $TiO_2$-terminated STO substrate, two unit cells of $SrRuO_3$ used to *switch* from $TiO_2$-termination to SrO-termination, and a cap of two unit cells of STO to verify the SrO-termination.[26,28] Note that the two unit cells of $SrRuO_3$ used to switch termination are actually two and a half unit cells, layered as $SrO/RuO_2/SrO/RuO_2/SrO$, as reported elsewhere.[37] We collected ten Auger spectra of each element, to improve the signal quality, nominally after every ¼$^{th}$ monolayer of growth, and the area under the curve were calculated from the sum of the spectra collected at each point. The details of the spectra and data processing can be found in the Experimental Section and Supplementary Information.

The resulting intensity values, starting with the spectra collected from the substrate pre-growth, are plotted as a function of the film thickness, shown in **Figure 2**. The Sr/Ti signal-ratio remains approximately constant, and consistent with the substrate, during the first two monolayers of deposition. During the subsequent deposition of two monolayers of $SrRuO_3$, the Sr and Ru signals increase while the Ti signal decreases exponentially, as would be expected with the corresponding increase/decrease in Sr, Ru, and Ti content relative to the surface of the film. The last two monolayers of deposited STO show a sustained higher level of Sr signal, decay of the Ru signal, and a recovery of the Ti signal, with the resulting Sr/Ti signal-ratio opposing that observed in $TiO_2$-terminated STO. This indicates that the termination was switched successfully from $TiO_2$ surface to SrO surface and maintained through the deposition of STO.

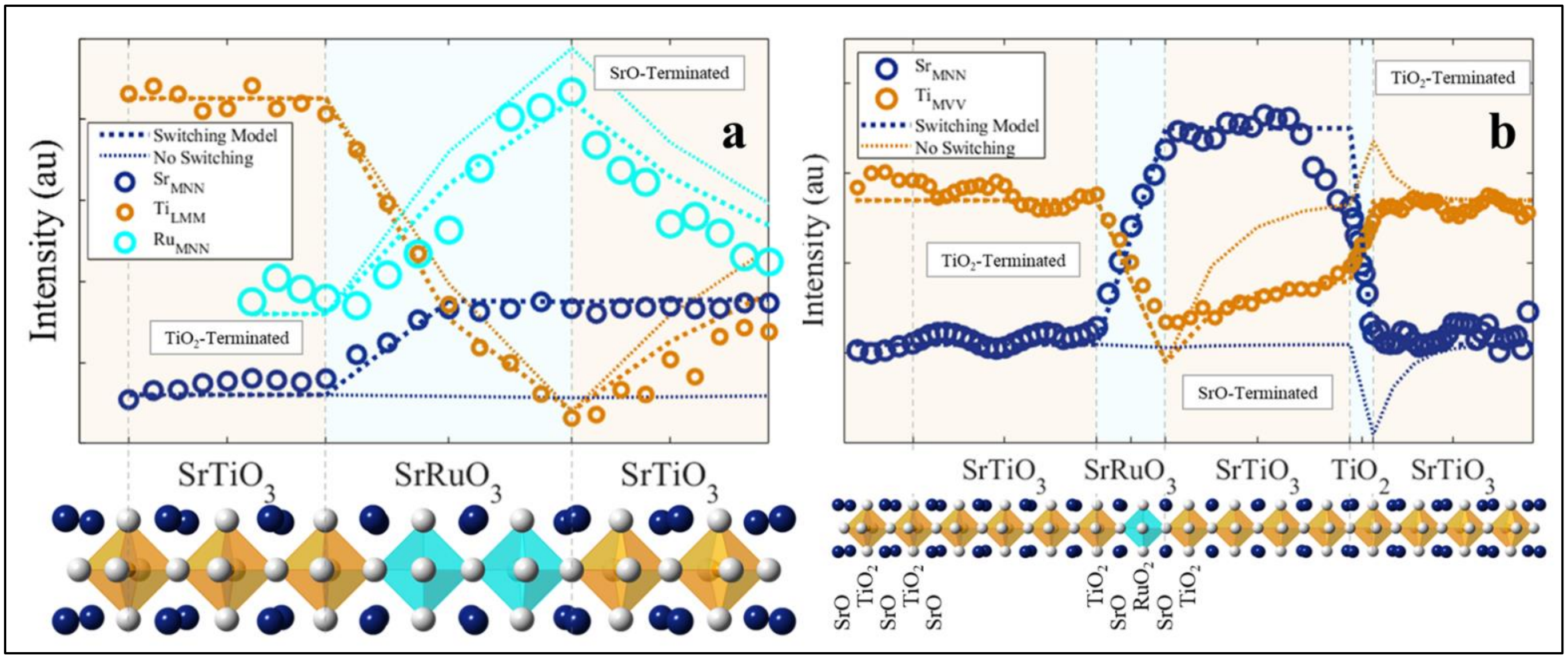

**Figure 2.** Auger electron spectra signal intensity calculated from areas beneath the curves for spectra collected (a) at known thickness intervals, and (b) in real time, during the deposition of homoepitaxial STO on a $TiO_2$-terminated STO substrate, with termination switching controlled by the deposition of $SrRuO_3$ or $TiO_2$. The Auger lines used to monitor Sr, Ru, and Ti are $Sr_{MNN}$, $Ru_{MNN}$, and $Ti_{LMM}$ transitions in (a), and to monitor Ti and Sr are the $Ti_{MVV}$ and $Sr_{MNN}$ transitions in (b), and their intensities clearly track the surface termination of the film both before and after all switching events. The dashed lines are modeled signal intensity for the structures, described in the text, with thick lines corresponding to the assumed switching events shown in the structure beneath the plot, and thin lines corresponding to the same growth without termination switching. Marker size is proportional to error.

We then demonstrate our ability to probe and control the surface termination switching *in situ* and real time using the pulse-probe technique. For this growth, we used $TiO_2$-terminated STO and grew, in chronological order, four unit cells of STO, one unit cell of $SrRuO_3$, four unit cells of STO, one monolayer of $TiO_2$, and eight unit cells of STO. The one unit cell of $SrRuO_3$ results in SrO-termination, and the termination is then switched back to $TiO_2$-termination with the deposition of a single monolayer of $TiO_2$. The area under the curve were calculated for each scan, as described in the Experimental Section and Supplementary Information, and their intensity as a function of film thickness is shown in Figure 2 (b). The relative signals of Sr and Ti follow the same trend as that observed in the heterostructure shown in Figure 2 (a) during the switch from the $TiO_2$-terminated substrate to SrO termination, before reversing to the original $TiO_2$ termination.

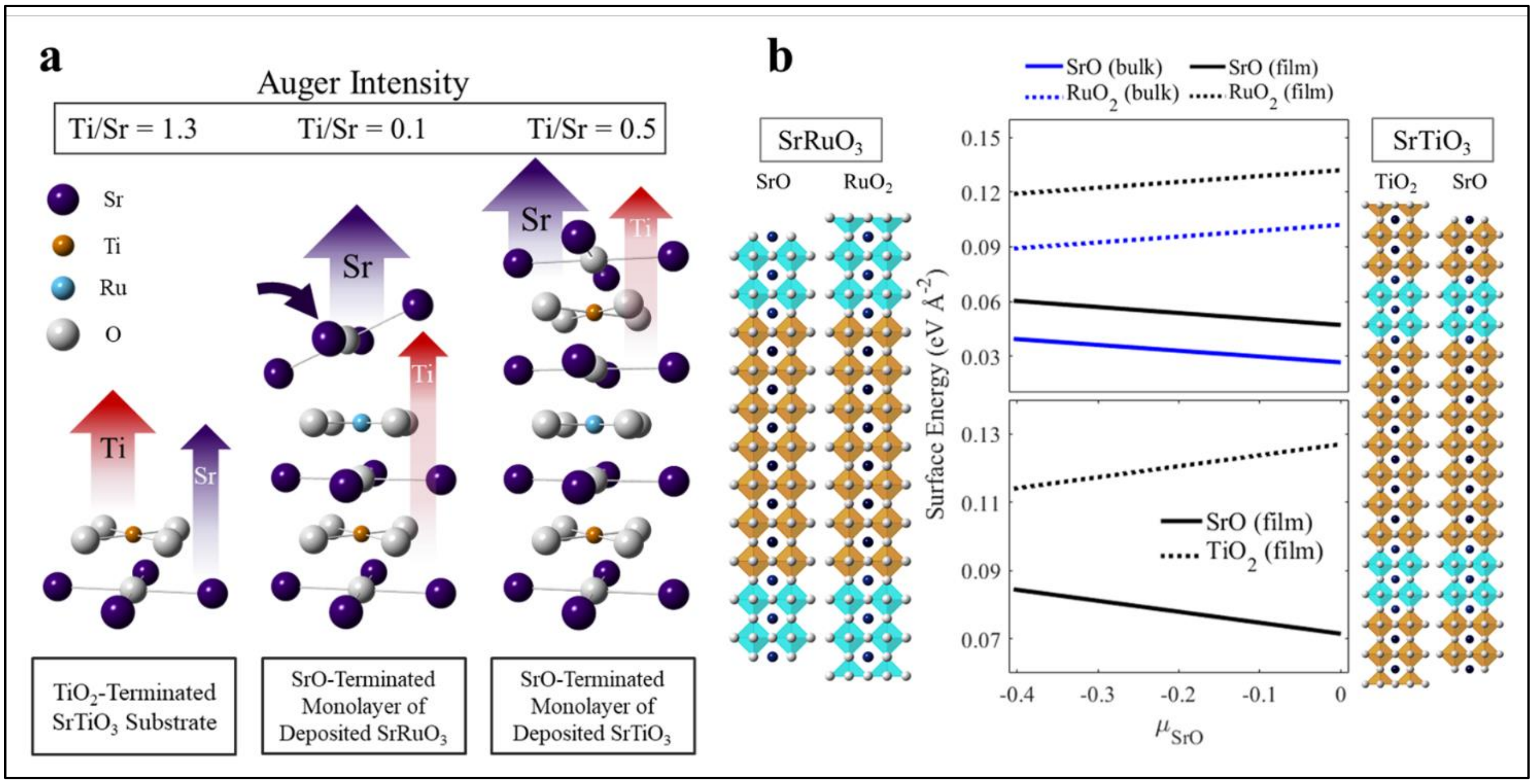

**Figure 3.** (a) Schematic depiction of the termination switching process in STO/SRO heterostructures, in which the deposition of SRO on $TiO_2$-terminated STO results in the formation of an "extra" layer of SrO, and thus SrO-termination which is maintained with additional deposition of STO. Arrows labeled as "Ti" or "Sr" illustrate relative Auger signal strength escaping the material after absorption by capping layers, forming the basis of the model used to approximate shifts in Auger signal with various heterostructures, quantified here as Ti/Sr. (b) Model configuration of SrO-terminated and $RuO_2$-terminated SRO films (left side) on an STO substrate, as well as $TiO_2$-terminated and SrO-terminated STO films sandwiching an SRO layer (right side).

Comparison of the chemical potential-dependent surface energy of the SRO- (top-center) and STO-capped (bottom-center) structures.

To quantitatively understand the termination changes in these heterostructures, we built a parameter-free model based on Auger electron escape depths, illustrated in **Figure 3**. The model accounts for variation in Auger signal due to compositional changes at the sub-monolayer scale by generating relative signal intensity shifts as a function of deposited layer thickness. It predicts the relative Auger electron intensities by examining how the signals from alternating layers in (100)-oriented STO, and epitaxial $SrRuO_3$, are attenuated by depth using calculated inelastic mean free paths (IMFPs) for the transitions observed.[39] Full details of the modeling methods can be found in the Supplementary Information. We compare the calculated relative AES intensities for each element with and without a termination switching event with the measured Auger data in Figure 2. The quality of the model's fits to the data is compared by calculating $\chi^2$, with the results normalized for comparison such that the non-switching $\chi^2 = 1$. While the switching versus non-switching models for the Ti and Ru signals appear similar in both the scenarios, as shown in Figure 2 (a), the rate of decay of Ti signal and its recovery during and after the deposition of $SrRuO_3$, and the inverse for Ru, has a better fit for the model with termination switching (for Ru $\chi^2 = 0.45$, and for Ti $\chi^2 = 0.30$). Likewise, the Sr signal follows the switching model exceedingly well, and demonstrates a deviation which cannot be explained without a termination switching event from $TiO_2$ to SrO (Sr $\chi^2 = 0.02$). The model with termination switching accurately predicts the evolution of the AES signal intensity including subtleties of the Ti signal recovery after the deposition of $SrRuO_3$, as shown in Figure 2 (b) (for Sr $\chi^2 = 0.02$, and for Ti $\chi^2 = 0.12$). The model depicting the signal from the same structure without the assumption of a switching event has an $RuO_2$- terminated $SrRuO_3$ monolayer, rather than SrO-terminated, and the deposition of $TiO_2$ results in a double-layer of $TiO_2$. The model without the switching event is clearly not a fit to the data for either Sr or Ti. The Ti signal without a switching event would increase above the initial intensity and then decay, which is not observed experimentally. The Sr signal would remain nearly constant after the deposition of $SrRuO_3$, with slight variation due to the shift in lattice constant, then decay sharply with the addition of another layer of $TiO_2$, neither of which were observed experimentally. This clearly shows, without ambiguity, that the measured AES data demonstrates the first real time and *in situ* observation and control of surface termination during complex oxide thin film deposition. The ability of such a simple model to accurately predict relative shifts in Auger signal using only IMFP and lattice dimensions is a testament to the sensitivity of this technique to subtle compositional changes on surfaces. Furthermore, the simplicity of both modeling and data collection will enable direct understanding of the growth mechanisms and compositional evolution during thin film growth, and engineering complex oxide heterostructures for a variety of applications.

We used DFT calculations to develop an understanding of the energetics of termination switching in STO/$SrRuO_3$ heterostructures observed experimentally. The details of the calculations are provided in the Supplementary Information. The results in Figure S4 show that in bulk STO, SrO-termination is more stable than $TiO_2$-termination for a broad range of chemical potentials, which is consistent with the literature.[43] In the case of bulk $SrRuO_3$ (Figure S5), however, SrO-termination is found to be more stable over the entire chemical potential range, where $SrRuO_3$ is stable. Moreover, when thin films of $SrRuO_3$ are deposited on STO substrates, the difference in energy between SrO-termination and $RuO_2$-termination increases over that of the bulk $SrRuO_3$, indicating that thin films tip the energetic preference even further in favor of SrO-termination. The stability difference between SrO- and $RuO_2$-termination in $SrRuO_3$ films can be understood from their electronic structures. The layer-resolved density of states of SrO-terminated $SrRuO_3$ on $SrTiO_3$ substrate (Figure S6) shows the presence of a lower density of states around the Fermi energy compared to $RuO_2$-termination (Figure S7), signifying the presence of a greater number of surface dangling bonds with $RuO_2$ termination, and thus its higher surface energy. Further deposition of additional layers of $SrTiO_3$ on a sandwiched layer of $SrRuO_3$ results in the stabilization of the SrO-termination layer in STO instead of the common $TiO_2$-termination. Our calculations in Figure 3(b) show that SrO-termination has a lower surface energy with reference to $TiO_2$-termination over the entire range of allowed chemical potential, in a manner similar to that of $SrRuO_3$ on STO substrates. It is consistent with our experimental observation that the termination layer is switched from $TiO_2$ to SrO for the STO epilayers deposited on $SrRuO_3$. The lower surface energy of SrO-terminated $SrTiO_3$ can be understood from the chemical potential constraints of $SrTiO_3$/$SrRuO_3$ hybrid structures. As mentioned above, the growth window for the formation of $SrRuO_3$ is narrower than that of $SrTiO_3$ (Figure S3). It is expected that the chemical potential constraints of $SrRuO_3$ will determine the growth condition of $SrRuO_3$/$SrTiO_3$ heterostructures. Due to the reduced growth window of $SrTiO_3$ on $SrRuO_3$, the SrO-terminated surface of $SrTiO_3$ has lower a surface energy in the whole allowed chemical potential range. It is different from that of pure bulk $SrTiO_3$, where a $TiO_2$-terminated surface could have a lower surface energy for a narrow chemical potential range (Figure S4).

## CONCLUSION

In summary, we have shown, for the first time, direct observation and control of surface termination in real time and *in situ* in PLD-grown perovskite oxide thin films using Auger electron spectroscopy. The evolution of Auger spectra in both static and real time studies agree well with a simple parameter-free model, validating this technique for surface analysis with monolayer resolution. The potential of this technique for observing subtle chemical and compositional dynamics of thin film growth methods is demonstrated, thereby providing a means for studying *in situ* events during growth, which have previously proven too challenging to monitor and control. The capability for monitoring real time growth events with atomic-level resolution is invaluable for continued advancement of thin film heterostructure engineering. With continued advances and deployment of machine learning approaches, combining real time growth monitoring with machine learning can lead to precision manufacturing of heterostructures for the next generation of electronic and photonic devices.

## EXPERIMENTAL SECTION

*Thin Film Growth*: Complex oxide thin films were grown using PLD with a 248 nm KrF laser operating at 1-5 Hz. Substrates were single crystal (001) $SrTiO_3$ purchased from CrysTec, then etched and annealed in flowing $O_2$ to achieve $TiO_2$-termination, as described elsewhere.[44] Growths were conducted at 800°C in $10^{-2}$ to $10^{-3}$ mbar $O_2$, with heating of the substrate performed under growth conditions. The $SrTiO_3$ target used was single-crystalline, also purchased from CrysTec, while the $SrRuO_3$ and $TiO_2$ targets were polycrystalline and made in-house. Ablation of the targets was conducted with a laser fluence of 0.7 – 1.8 J cm$^{-2}$ with a focused spot approximately 2.5 mm$^2$. Thickness monitoring was performed with RHEED, utilizing an electron source operating with a 5 kV accelerating voltage and 5 μA emission current.

*Auger Electron Spectroscopy*: Auger spectra were collected with a Staib AugerProbe™ and its accompanying software suite. The same electron source was used for generating the Auger electrons as used for RHEED, operating with the same parameters. The transitions selected for observation for the demonstration of the pulse-probe technique, shown in Figure 3, were $O_{KLL}$, $Ti_{LMM}$, and $Sr_{LMM}$ located at approximately 518, 382, and 1670 eV, respectively. Auger spectra were collected for energy ranges of 2 eV between pulses, with peak locations optimized using spectra collected from the STO substrate. The resulting signal is the intensity of the peaks, with details and examples shown in the Supplementary Information. The transitions selected for the observation of termination switching, shown in Figure 1, were $O_{KLL}$, $Ti_{LMM}$, and $Sr_{MNN}$. We acquired ten scans of the Auger spectra for each element to improve signal quality, and area under the curve values were calculated from the sum of the spectra collected at each point, as explained further in the Supplementary Information. To observe termination switching both in real time and *in situ*, as shown in Figure 2, we selected the transitions with the lowest IMFP while still maintaining sufficient intensity. For STO this meant the highest probability $Sr_{MNN}$ and $Ti_{MVV}$ transitions at approximately 85 and 35 eV, respectively. We optimized the scan widths and rates to provide the highest signal-to-noise with minimal scan duration, and a pulse-to-scan ratio of 3:1 was selected to allow monitoring of RHEED oscillations despite the long recovery time between pulses which was, in this case, approximately 20 seconds. RHEED oscillations were quite clear, making *in situ* thickness determination trivial, as shown in the supplemental information.

*Density-Functional-Theory Calculations*: We used the Vienna Ab initio Simulation Package to carry out the DFT calculations.[45] The energy cutoff for the plane waves was set at 500 eV. The threshold for energy convergence of the self-consistent loops was set to $10^{-6}$ eV. During structure optimization, the convergence criteria for forces on ions was set to 0.01 eV $Å^{-1}$. We used projector augmented-wave potentials and the generalized gradient approximation within the Perdew-Burke-Ernzerhof parameterization to describe the electron-ion and the electronic exchange-correlation interactions, respectively.[46,47] Additional details may be found in the supporting information.

## ASSOCIATED CONTENT

Example Auger Spectra, Modeling Auger Signal, Pulse-Probe RHEED Oscillations, Surface Energy Calculations

## AUTHOR INFORMATION

### Corresponding Author

*j.ravichandran@usc.edu

### Present Addresses

⸙ Core Center for Excellence in Nano Imaging, University of Southern California, 925 Bloom Walk, Los Angeles, CA 90089, USA

† Department of Mechanical Engineering & Materials Science, and Institute of Materials Science & Engineering, Washington University in St. Louis, One Brookings Drive, St. Louis, MO 63130, USA

§ Ming Hsieh Department of Electrical and Computer Engineering, University of Southern California, 925 Bloom Walk, Los Angeles, CA 90089, USA

### Author Contributions

The manuscript was written through contributions of all authors. All authors have given approval to the final version of the manuscript.

**Funding Sources**

This work was supported by the Air Force Office of Scientific Research under contract FA9550-16-1-0335 and Army Research Office under Award No. W911NF-19-1-0137. The work at Washington University was supported by the National Science Foundation through grant number DMR-1806147. This work used the computational resources of the Extreme Science and Engineering Discovery Environment (XSEDE), which is supported by NSF ACI-1548562.

**ACKNOWLEDGMENT**

T.O and M.S acknowledge the Andrew and Erna Viterbi Graduate Student Fellowship, and H. K. acknowledges the Annenberg Graduate Student Fellowship. The authors gratefully acknowledge support from Staib Instruments, as well as the benefit of discussions with Dr. Philippe G. Staib, Dr. Eric Dombrowski, and Laws Calley.